\documentclass[11pt,a4paper]{article}
\usepackage[T1]{fontenc}
\usepackage{lmodern}
\usepackage[margin=2.5cm]{geometry}
\usepackage{amsmath,amssymb,amsthm,mathtools}
\usepackage{graphicx,booktabs,array}
\usepackage[numbers,sort&compress]{natbib}
\usepackage[hidelinks]{hyperref}

\newtheorem{proposition}{Proposition}
\newtheorem{corollary}[proposition]{Corollary}
\newtheorem{theorem}[proposition]{Theorem}
\theoremstyle{remark}
\newtheorem{remark}{Remark}
\allowdisplaybreaks
\newcommand{\E}{\mathbb{E}}
\newcommand{\Z}{\mathbb{Z}}
\newcommand{\one}{\mathbf{1}}
\newcommand{\cH}{\mathcal{H}}

\title{Exact structure and metastability in a probabilistic cellular automaton between Wolfram rules 23, 77, 178 and 232}
\author{Francisco J. Mu\~noz and Juan Carlos Nu\~no\thanks{Corresponding author: \href{mailto:juancarlos.nuno@upm.es}{juancarlos.nuno@upm.es}}\\[4pt]
\small Departamento de Matem\'atica Aplicada, Universidad Polit\'ecnica de Madrid, 28040 Madrid, Spain}
\date{}

\begin{document}
\maketitle

\begin{abstract}
We study a one-dimensional probabilistic cellular automaton that interpolates between Wolfram's elementary rules 23, 77, 178 and 232. Each cell follows its two neighbours with one probability when they agree and keeps its own state with a second probability when they disagree, so the cell itself acts only as a tie-breaker; the four deterministic rules sit at the corners of the parameter square and the noisy majority vote lies on its diagonal. Earlier exact results were limited to rings of at most four cells. We obtain two main results valid for every ring size. First, when ties are broken by a fair coin the model is exactly solvable: the stationary state is an Ising chain along the two interleaved sublattices of the ring, the spectral gap does not depend on the size, and the correlation length and relaxation time diverge only at the boundary of the square, with diffusive scaling, so that the edges behave as zero-noise critical points. Second, the discontinuities at the edges reported earlier are replaced, at small distance from the edge, by a metastable time scale inversely proportional to that distance, with a prefactor obtained in closed form for even sizes through a mapping onto compact directed percolation. Both results rest on the decomposition of the ring into interleaved sublattices, which also selects the singular edges for each size. Exact symmetries, matrix-free linear algebra and phase-type methods extend the analysis to rings of twenty-two cells and show that consensus times grow exponentially with size.

\end{abstract}

\section{Introduction}\label{sec:intro}

Probabilistic cellular automata (PCA) are discrete-time Markov chains on configuration spaces in which all cells are updated synchronously and independently given the current configuration \cite{MairesseMarcovici2014}. They are minimal models of noisy collective dynamics and of opinion formation, and a classical testing ground for the question of whether local noise always destroys long-range order \cite{Gacs2001}. In \cite{MunozNuno2026} we introduced a one-dimensional binary PCA with nearest-neighbour interactions that interpolates between the elementary rules 23, 77, 178 and 232 of Wolfram's classification \cite{Wolfram1983}. The parameter $r$ is the probability of adopting the state of two agreeing neighbours, and $p$ is the probability of keeping the current state when they disagree. We solved the model exactly for rings of $L=2,3,4$ cells. In the interior of the square, $0<p,r<1$, the stationary distribution is unique; on the edges and at the vertices, eigenvalues of modulus one appear, the asymptotic behaviour depends on the initial condition, and some behaviours, such as period-two oscillations between the two consensus states, are not the limit of the interior behaviour. The exponential growth of the transition matrix prevented us from going further.

In this paper we show that the model has an exact structure, valid for every ring size, that explains these observations and makes larger rings accessible. Three ideas organise the paper.

\paragraph{Stride-two cycles.} When the two neighbours of cell $i$ are $i-1$ and $i+1$, it is natural to read the ring along the \emph{stride-two cycles} $i\mapsto i+2$: two cycles of length $L/2$ (the even and the odd cells) for even $L$, and a single cycle of length $L$ for odd $L$. This decomposition turns out to control the whole model. On the line $p=1/2$, where the central cell is ignored, the dynamics decouples along these cycles and becomes exactly solvable (Section~\ref{sec:half}). On the edges $r\in\{0,1\}$, the space-time version of the same cycles is conserved, which selects the absorbing states according to the parity of $L$ (Section~\ref{sec:edges}). On the edges $p\in\{0,1\}$, frozen configurations are alternating patterns along the cycles, which close only when $4\mid L$.

\paragraph{The edges are zero-noise critical points.} At $p=1/2$ the stationary state is an Ising chain along each stride-two cycle with an explicit coupling. Its correlation length $\ell$ is finite in the interior and diverges only at the edges, where, at distance $\varepsilon$, $\ell\simeq1/(2\sqrt\varepsilon)$ while the relaxation time is $1/(2\varepsilon)\simeq2\ell^2$. The edges thus behave as zero-temperature critical points with diffusive dynamics, as the one-dimensional Ising chain under Glauber dynamics \cite{Glauber1963}. Neither exact results nor numerics show any sign of a phase transition inside the square (Section~\ref{sec:infinite}).

\paragraph{Metastability replaces the discontinuities.} The discontinuities found in \cite{MunozNuno2026} are not abrupt changes of the stationary state at finite $L$, which is a rational function of $(p,r)$ in the interior, but a non-commutativity of limits. Close to an edge the chain spends a long time near the absorbing states of the edge, and leaves them on a metastable time scale $t^*\simeq1/(\mu\varepsilon)$. The rate $\mu$ is given by absorption probabilities of a phase-type (PH) problem \cite{Neuts1981,LatoucheRamaswami1999} and, for even $L$, is obtained in closed form through a mapping onto compact directed percolation \cite{DomanyKinzel1984,Essam1989} (Section~\ref{sec:meta}).

\medskip
Accordingly, the paper is built around two main results, both proved for every $L$:
\begin{itemize}
\item[\textbf{A.}] \emph{Exact solution on $p=1/2$} (Proposition~\ref{prop:half}, Theorem~\ref{thm:ising} and Corollary~\ref{cor:half}): spectrum, stationary distribution, correlations and consensus probabilities for every ring size, and the diffusive divergence at the edges.
\item[\textbf{B.}] \emph{Metastable time scale at the edges} (Proposition~\ref{prop:pert} and Theorem~\ref{thm:mu-even}): $t^*\simeq1/(\mu\varepsilon)$ with $\mu$ given in closed form for even $L$.
\end{itemize}
The symmetries of Section~\ref{sec:sym} (lumpability and two dualities) and the structure of the deterministic edges (Section~\ref{sec:edges}) are the tools behind them. The computational framework of Section~\ref{sec:comp} (lumping, matrix-free products and PH methods) extends the analysis to $L=22$; it serves to test the theory and to explore the rest of the square, and none of the results above depends on it. A few statements established by exact computation for finitely many sizes are collected in Appendix~\ref{app:verif}, and explicit closed forms and full numerical tables are given in Appendices~\ref{app:exact} and~\ref{app:tables}.

\section{Model and preliminaries}\label{sec:model}

We consider a ring of $L\ge 3$ cells with states $x_i\in\{0,1\}$, $i\in\Z_L$, updated synchronously. Given the current configuration $x$, the new states $y_i$ are conditionally independent with
\begin{equation}\label{eq:rule}
P(y_i=1\mid x)=\begin{cases} r & x_{i-1}=x_{i+1}=1,\\ 1-r & x_{i-1}=x_{i+1}=0,\\ p & x_{i-1}\neq x_{i+1},\ x_i=1,\\ 1-p & x_{i-1}\neq x_{i+1},\ x_i=0,\end{cases}
\end{equation}
which is Table~2 of \cite{MunozNuno2026}. The transition matrix is
\begin{equation}\label{eq:A}
A(x,y)=\prod_{i\in\Z_L}K\big(x_{i-1},x_i,x_{i+1};y_i\big),
\end{equation}
where $K(a,b,c;1)=P(y_i=1\mid a,b,c)$ is given by \eqref{eq:rule}. We write $0^L$ and $1^L$ for the two consensus configurations, $\cH=\{0^L,1^L\}$, and $\sigma_i=2x_i-1\in\{-1,1\}$. A direct case check of \eqref{eq:rule} gives the conditional mean
\begin{equation}\label{eq:mean}
\E[\sigma_i'\mid x]=\frac{2r-1}{2}\,(\sigma_{i-1}+\sigma_{i+1})+\frac{2p-1}{2}\,\sigma_i\,(1-\sigma_{i-1}\sigma_{i+1}),
\end{equation}
where the first term acts when the neighbours agree and the second when they disagree.

\paragraph{What makes the model special.} In a three-cell neighbourhood the majority is decided by the two outer cells whenever they agree; only when they disagree is the central cell decisive, as a tie-breaker. Rule 232 is the majority vote of the three cells, rule 23 its complement, and rules 178 and 77 follow and oppose, respectively, two agreeing neighbours while flipping, respectively keeping, the central cell at a tie. The PCA therefore separates two behavioural parameters that a noisy majority vote ties together: $r$ measures conformity to a local consensus, and $p$ measures stubbornness when the neighbourhood is split. The majority vote with independent output errors of probability $\eta$ is the diagonal $p=r=1-\eta$, and rule 178 with errors is the anti-diagonal $p=1-r$. On $p=1/2$ the central cell is ignored and the model becomes a two-input PCA. On $r\in\{0,1\}$ agreeing neighbourhoods are deterministic, and on $p\in\{0,1\}$ ties are deterministic. In the region $r\ge\max(p,1-p)$ the probability $P(y_i=1\mid x)$ is nondecreasing in $x$, so the PCA is attractive (monotone), as the noisy majority vote is.

\section{Symmetries and lumpability}\label{sec:sym}

\begin{proposition}[Lumpability]\label{prop:lump}
Let $G=D_L\times\Z_2$ act on $\{0,1\}^{\Z_L}$ by rotations, reflection $x_i\mapsto x_{-i}$ and global complementation $C:x\mapsto\bar x$. Then $A(gx,gy)=A(x,y)$ for all $g\in G$. Consequently, the partition into $G$-orbits is strongly lumpable \cite{KemenySnell1960}: for every pair of orbits $O,O'$ the quantity $B(O,O')=\sum_{y\in O'}A(x,y)$ does not depend on $x\in O$. For $0<p,r<1$ the stationary distribution $\pi$ is $G$-invariant, and the law of the hitting time of any $G$-invariant set from a $G$-invariant initial law is the same in the lumped chain.
\end{proposition}
\begin{proof}
Invariance under rotations is evident from \eqref{eq:A}. Reflection exchanges $x_{i-1}$ and $x_{i+1}$, and \eqref{eq:rule} is symmetric in them. Complementation maps every case of \eqref{eq:rule} to the complementary case, and $K(\bar a,\bar b,\bar c;\bar y)=K(a,b,c;y)$. The remaining statements are standard consequences: if $\pi$ is the unique stationary law then so is $\pi\circ g$.
\end{proof}

The number of orbits is close to $2^L/(4L)$; for instance it is $122$, $1162$ and $48\,734$ for $L=12,16,22$.

\begin{proposition}[Duality]\label{prop:dual}
Let $C$ also denote the permutation matrix of global complementation. Then $CA(p,r)=A(p,r)C$ and
\begin{equation}\label{eq:dual}
A(p,r)\,C=A(1-p,1-r).
\end{equation}
Hence $A(p,r)$ and $A(1-p,1-r)$ have the same stationary distribution in the interior, and their spectra coincide up to sign: eigenvalues of $C$-even eigenvectors are preserved and eigenvalues of $C$-odd eigenvectors change sign.
\end{proposition}
\begin{proof}
$(A C)(x,y)=A(x,\bar y)=\prod_i K(x_{i-1},x_i,x_{i+1};1-y_i)$. Every entry of \eqref{eq:rule} is mapped to one minus itself, which is the same table with $(p,r)$ replaced by $(1-p,1-r)$. Commutation follows from Proposition~\ref{prop:lump}. If $vA=\lambda v$ and $vC=\pm v$, then $vA(1-p,1-r)=vAC=\pm\lambda v$.
\end{proof}

The duality maps the vertices $23\leftrightarrow 232$ and $77\leftrightarrow178$, the edge $r=0$ onto $r=1$ (with $p\mapsto 1-p$) and the edge $p=0$ onto $p=1$ (with $r\mapsto 1-r$). In particular, the period-two oscillations between $0^L$ and $1^L$ on the edge $r=0$ are the image of the two absorbing consensus states of the edge $r=1$ seen through a complementation at every step.

\begin{proposition}[Staggered symmetry, even $L$]\label{prop:stag}
Let $L$ be even and let $S$ be the staggered complementation $x_i\mapsto x_i\oplus(i\bmod 2)$. Then $S\,A(p,r)\,S=A(p,1-r)$. Consequently $A(1-p,r)=S\,A(p,r)\,C\,S$, and for even $L$ the whole square is determined, up to signs of eigenvalues, by the quadrant $p,r\le 1/2$.
\end{proposition}
\begin{proof}
For even $L$, $S$ is well defined on the ring. For a cell with $i$ even, $x_{i\pm1}$ are complemented and $x_i,y_i$ are not: ``neighbours agree on $1$'' becomes ``neighbours agree on $0$'', so $r$ becomes $1-r$, and the tie case is unchanged. For $i$ odd, $x_i$ and $y_i$ are complemented and the neighbours are not: $P(\bar y_i=1\mid x)=1-P(y_i=1\mid x)$ again replaces $r$ by $1-r$ in the agreement case, while keeping $\bar x_i$ at a tie is the same event as keeping $x_i$, so $p$ is unchanged. The last statements follow by combining with Proposition~\ref{prop:dual}.
\end{proof}

For odd $L$ the staggered map does not exist (the ring is frustrated), and $r\mapsto 1-r$ is not a symmetry; already for $L=5,7$ the matrices $SA(p,r)S$ and $A(p,1-r)$ differ. This explains several differences between $L=3$ and $L=4$ in \cite{MunozNuno2026}.

\section{Main result A: the exactly solvable line \texorpdfstring{$p=1/2$}{p=1/2}}\label{sec:half}

Let $u=2r-1$. For $p=1/2$, Eq.~\eqref{eq:mean} reduces to the linear relation $\E[\sigma_i'\mid x]=\tfrac u2(\sigma_{i-1}+\sigma_{i+1})$.

\begin{proposition}\label{prop:half}
Let $p=1/2$ and $L\ge3$. Then:
\begin{enumerate}
\item[(a)] The eigenvalues of $A$ are $\bigcup_{n=0}^{L}u^n\,\mathrm{spec}(M_n)$, where $M_n$ is a nonnegative substochastic matrix indexed by $n$-subsets of $\Z_L$ (defined in the proof). In particular $M_1=\tfrac12(\mathcal U+\mathcal U^{-1})$, where $\mathcal U$ is the rotation $i\mapsto i+1$, so $\lambda_{1,k}=u\cos(2\pi k/L)$, $k=0,\dots,L-1$.
\item[(b)] The second largest eigenvalue modulus is $|\lambda_2|=|2r-1|$ for every $L$. The relaxation time is $t_{\rm rel}=1/(1-|\lambda_2|)=1/(2\min(r,1-r))$, independent of $L$.
\item[(c)] The magnetization satisfies $\E[\sum_i\sigma_i(t)]=u^t\,\E[\sum_i\sigma_i(0)]$ exactly.
\end{enumerate}
\end{proposition}
\begin{proof}
Let $A$ act on functions by $(Af)(x)=\E[f(y)\mid x]$. By conditional independence, for $S\subset\Z_L$ with $|S|=n$,
\[
A\sigma_S=\Big(\frac u2\Big)^{n}\prod_{i\in S}(\sigma_{i-1}+\sigma_{i+1}),\qquad \sigma_S=\prod_{i\in S}\sigma_i .
\]
Expanding the product gives $2^n$ terms. Each term is the monomial $\sigma_{\phi(S)}$ when the neighbour choice $\phi(i)\in\{i-1,i+1\}$ is injective; otherwise $\sigma_j^2=1$ lowers the degree. Thus $V_n=\mathrm{span}\{\sigma_S:|S|\le n\}$ is invariant, and on $V_n/V_{n-1}$ the operator acts as $u^nM_n$, with $M_n(S,S')=2^{-n}\#\{\phi\ \text{injective}:\phi(S)=S'\}$. Its rows have nonnegative entries summing to at most one, so $\rho(M_n)\le1$. The spectrum of $A$ is the union of the spectra of the diagonal blocks, which proves (a). Then (b) follows because $\lambda=u$ ($n=1$, $k=0$) attains $|u|$ while every eigenvalue with $n\ge2$ has modulus at most $u^2\le|u|$. Part (c) is (a) applied to the eigenfunction $\sum_i\sigma_i$.
\end{proof}

Order the cells along the stride-two cycles of the ring: for even $L$ the even and the odd cells, each a cycle of length $n=L/2$; for odd $L$ the single cycle $(0,2,4,\dots,L-1,1,3,\dots,L-2)$ of length $n=L$. For $\zeta\in[0,1)$, the \emph{Ising (Markov) measure} on a cycle $(c_1,\dots,c_n)$ is
\begin{equation}\label{eq:ising}
\nu_\zeta(\sigma)=\frac{1}{2^n(1+\zeta^n)}\prod_{k=1}^{n}\big(1+\zeta\,\sigma_{c_k}\sigma_{c_{k+1}}\big),\qquad c_{n+1}=c_1 ,
\end{equation}
that is, a stationary symmetric Markov chain along the cycle with probability $(1+\zeta)/2$ of keeping the same state, conditioned to close.

\begin{theorem}[Stationary distribution at $p=1/2$]\label{thm:ising}
Let $p=1/2$, $0<r<1$, $u=2r-1$, and
\begin{equation}\label{eq:zeta}
\zeta=\xi^2,\qquad \xi=\frac{1-\sqrt{1-u^2}}{|u|}\quad(\xi=0\text{ if }u=0),
\end{equation}
which is the unique root in $[0,1)$ of $\zeta=\tfrac{u^2}{4}(1+\zeta)^2$. Then the stationary distribution of the PCA on the ring of size $L\ge3$ is the product of the Ising measures $\nu_\zeta$ over the stride-two cycles: $\pi=\nu_\zeta\otimes\nu_\zeta$ (one factor per parity class) for even $L$, and $\pi=\nu_\zeta$ on the single cycle of length $L$ for odd $L$.
\end{theorem}
\begin{proof}
At $p=1/2$ a cell whose neighbours disagree takes a fair coin, independently of everything else, and a cell whose neighbours agree copies them with probability $r$. Hence $y_i$ depends only on the pair $(x_{i-1},x_{i+1})$ and on independent noise. The two neighbours of $i$ are consecutive on a stride-two cycle, and so are the cells $i$ and $i+2$ that they produce. Therefore the new configuration along a cycle is the output of a hidden Markov model whose hidden sequence is the old configuration along a cycle and whose emissions depend on consecutive hidden pairs. For even $L$ the two parity classes are updated from disjoint inputs with independent noise, so independence between them is preserved, and it suffices to show that $\nu_\zeta$ is mapped to itself.

If the hidden sequence has law \eqref{eq:ising}, its transfer matrix is $\mathcal T=\begin{psmallmatrix}s&1-s\\1-s&s\end{psmallmatrix}$, $s=(1+\zeta)/2$, and the probability of an output word $y$ on the cycle is $\mathrm{Tr}\prod_k M_{y_k}/\mathrm{Tr}\,\mathcal T^n$, where $M_y(a,c)=\mathcal T(a,c)\,P(y\mid a,c)$ and $P(y\mid a,c)$ is $1/2$ if $a\ne c$, $r$ if $y=a=c$ and $1-r$ if $y\ne a=c$. In the Hadamard basis $H=\tfrac1{\sqrt2}\begin{psmallmatrix}1&1\\1&-1\end{psmallmatrix}$, using $M_1+M_0=\mathcal T$ and $M_1-M_0=su\,\mathrm{diag}(-1,1)$,
\[
HM_yH=\tfrac12\,D(\varepsilon)\,N_0\,D(\varepsilon),\qquad N_0=\begin{pmatrix}1&-su\\-su&\zeta\end{pmatrix},\quad D(\varepsilon)=\mathrm{diag}(1,\varepsilon),
\]
with $\varepsilon=+1$ for $y=1$ and $\varepsilon=-1$ for $y=0$. Since $D(\varepsilon_k)D(\varepsilon_{k+1})=D(\eta_k)$ with $\eta_k=\varepsilon_k\varepsilon_{k+1}$ (indices modulo $n$), cyclicity of the trace gives $\mathrm{Tr}\prod_kM_{y_k}=2^{-n}\mathrm{Tr}\prod_k N_0D(\eta_k)$. The fixed-point equation $\zeta=s^2u^2$ is exactly the condition $\det N_0=0$, so $N_0=vv^{\top}$ with $v=(1,-su)^{\top}$, and
\[
\mathrm{Tr}\prod_k N_0D(\eta_k)=\prod_k v^{\top}D(\eta_k)v=\prod_k\big(1+\zeta\,\eta_k\big).
\]
Dividing by $\mathrm{Tr}\,\mathcal T^n=1+\zeta^n$ gives exactly $\nu_\zeta(y)$ in \eqref{eq:ising}. Hence the product measure in the statement is stationary, and it is the stationary distribution because the chain is irreducible and aperiodic for $0<r<1$.
\end{proof}

\begin{remark}[A probabilistic reading of the proof]\label{rem:prob}
Since $\det M_1=\det M_0$ by symmetry, the condition $\det N_0=0$ says that both matrices $M_y$ have rank one. As $\tfrac12M_y(a,c)$ is the joint probability that two consecutive old cells take the values $a,c$ and produce the new value $y$, rank one means that \emph{the two old cells are conditionally independent given the new cell they produce}. Consequently, given $y_k$, the old cell shared by $y_k$ and $y_{k+1}$ is independent of everything produced before, and the new configuration is again Markov along the cycle. Its nearest-neighbour correlation follows from \eqref{eq:mean}: two consecutive new cells along a cycle come from three consecutive old cells, so $\E[\sigma'\sigma']=\tfrac{u^2}{4}\E[(\sigma_1+\sigma_2)(\sigma_2+\sigma_3)]=\tfrac{u^2}{4}(1+\zeta)^2$ up to finite-size corrections, and requiring it to equal $\zeta$ gives the same equation for $\zeta$. The fixed point is thus selected twice, by Markovianity and by self-consistency of the correlation.
\end{remark}

\begin{remark}[Relation to two-input PCA]
After relabelling along the stride-two cycles, the PCA at $p=1/2$ is a PCA with two inputs, $P(y=1\mid a,c)=\theta_{ac}$ with $\theta_{11}=r$, $\theta_{00}=1-r$ and $\theta_{01}=\theta_{10}=1/2$. Two-input binary PCA with a Markov invariant measure were characterised on $\Z$ by Belyaev, Gromak and Malyshev \cite{BGM1969} and on the periodic line by Bousquet-M\'elou \cite[Prop.~4.6]{BousquetMelou1998}, through the condition $\theta_{00}\theta_{11}(1-\theta_{01})(1-\theta_{10})=\theta_{01}\theta_{10}(1-\theta_{00})(1-\theta_{11})$; see also \cite{CasseMarckert2015,MairesseMarcovici2014}. Here both sides equal $r(1-r)/4$, so Theorem~\ref{thm:ising} is consistent with, and could be derived from, those results. The proof above is self-contained and gives $\zeta$ explicitly; what is specific to the present model is the reduction of a three-input rule to a two-input one on the stride-two cycles, whose number and length depend on the parity of $L$.
\end{remark}

\begin{corollary}\label{cor:half}
Let $p=1/2$ and $0<r<1$.
\begin{enumerate}
\item[(a)] For even $L$, $C_d=\E_\pi[\sigma_0\sigma_d]=0$ for odd $d$ and $C_{2m}=(\zeta^m+\zeta^{n-m})/(1+\zeta^{n})$, $n=L/2$. For odd $L$, $C_d=(\zeta^k+\zeta^{L-k})/(1+\zeta^L)$, where $2k\equiv d\pmod L$. As $L\to\infty$, $C_d\to\xi^d$ for even $d$ and $C_d\to0$ for odd $d$.
\item[(b)] For even $L$ the density of domain walls and the probability that a cell changes state in one step are exactly $1/2$.
\item[(c)] The consensus weight is $\pi(0^L)=\big[(1+\zeta)^n/(2^n(1+\zeta^n))\big]^{2}$ with $n=L/2$ for even $L$, and $\pi(0^L)=(1+\zeta)^L/(2^L(1+\zeta^L))$ for odd $L$. In particular $\pi(0^L)\sim\theta^L$ with $\theta=(1+\zeta)/2$.
\item[(d)] On $\Z$, the product of two independent stationary symmetric Markov chains on the even and odd sublattices, each with correlation $\zeta$ between consecutive sites of the sublattice, is an invariant measure of the PCA.
\item[(e)] The correlation length $\ell=-1/\ln\xi$ is finite for $0<r<1$. As $r=1-\varepsilon\to1$ (and likewise as $r=\varepsilon\to0$),
\begin{equation}\label{eq:diffusive}
\ell\simeq\frac{1}{2\sqrt\varepsilon},\qquad t_{\rm rel}=\frac{1}{2\varepsilon}\simeq2\,\ell^{2}.
\end{equation}
\end{enumerate}
\end{corollary}
\begin{proof}
(a) and (c) are standard properties of \eqref{eq:ising}. For (b), the wall density is $(1-C_1)/2$ and the flip probability is $(1-\E[\sigma_i\sigma_i'])/2=(1-uC_1)/2$ by \eqref{eq:mean}; both equal $1/2$ because $C_1=0$. (d) follows from the same transfer-matrix computation on $\Z$, with the trace replaced by the stationary boundary vector $(1,1)/\sqrt2$, which is the first Hadamard basis vector; we do not claim uniqueness. For (e), $u=1-2\varepsilon$ gives $\sqrt{1-u^2}=2\sqrt\varepsilon+O(\varepsilon^{3/2})$ and $\xi=1-2\sqrt\varepsilon+O(\varepsilon)$, and $t_{\rm rel}$ is Proposition~\ref{prop:half}(b).
\end{proof}

Equation~\eqref{eq:diffusive} identifies the edges $r\in\{0,1\}$ as zero-noise critical points with dynamic exponent $z=2$, $t_{\rm rel}\propto\ell^{z}$, the same diffusive scaling as the one-dimensional Ising chain with Glauber dynamics at zero temperature \cite{Glauber1963}: domain walls perform random walks and annihilate, and there is no ordered phase at positive noise. Proposition~\ref{prop:half}(c) also shows the mechanism behind the discontinuity at $r=0$ in its simplest form: for small $r$ the magnetization oscillates with factor $u\approx-1$, and the oscillation is damped only on the time scale $1/(2r)$. All formulas agree with the numerically computed stationary distributions to $10^{-12}$ for $3\le L\le22$; for example, at $r=0.9$, $\zeta=1/4$ and Corollary~\ref{cor:half}(c) gives $\pi(0^{20})=8.271790\ldots\times10^{-5}$, as in Table~\ref{tab:summary}.

\section{Computational framework}\label{sec:comp}

Away from $p=1/2$ we rely on four tools.

\emph{Lumping.} By Proposition~\ref{prop:lump} the chain can be lumped onto $G$-orbits, which reduces the state space by a factor of about $4L$. Combined with exact rational arithmetic, this gives $\pi$ and hitting-time means as exact fractions for $L\le8$. It reproduces the closed forms of \cite{MunozNuno2026} for $L=3,4$, and exact values at rational points can be turned into bivariate closed forms in $(p,r)$ for $L\le5$ (Appendix~\ref{app:verif-pi}; exact values and the formulas are listed in Appendix~\ref{app:exact}).

\emph{Matrix-free products.} Although $A$ is dense in the interior, \eqref{eq:A} is a product of local factors, so $w=vA$ can be computed by a sweep: $v$ is treated as a tensor with $L$ binary indices, the old bit $x_i$ is replaced by the new bit $y_i$ site by site, and $x_i$ is summed out as soon as no later site needs it. At most $L+3$ binary indices are alive at any time, so the cost is $O(L\,2^L)$ operations with $O(2^L)$ memory (about $3$\,s per product for $L=22$ on one core); $Am$ is computed in the same way. Stationary distributions follow by power iteration, and the leading eigenvalues by implicitly restarted Arnoldi on the linear operator \cite{ARPACK}.

\emph{Phase-type hitting times.} Let $\tau_c$ be the first time the PCA hits $\cH$ from an initial law $\alpha$, and let $T$ be the restriction of $A$ to $\cH^c$. Then $\tau_c$ has a discrete PH distribution with representation $(\alpha,T)$ \cite{Neuts1981,LatoucheRamaswami1999}: $P(\tau_c>k)=\alpha T^k\one$ and $\E[\tau_c]=\alpha(I-T)^{-1}\one$. The Perron root $\rho(T)<1$ gives the geometric tail $P(\tau_c>k)\sim c\,\rho(T)^k$, and its left eigenvector is the quasi-stationary distribution \cite{DarrochSeneta1965}. For large $L$ we propagate $w_k=\alpha T^k$ with the matrix-free product, setting the coordinates in $\cH$ to zero after each step. The survival ratios $S_{k+1}/S_k$ converge to $\rho(T)$, and $\E[\tau_c]=\sum_{k<K}S_k+S_K\rho(T)/(1-\rho(T))$ once $K$ is in the geometric regime ($30$--$70$ products). The same PH formalism, with absorption instead of survival, gives the metastable rates of Section~\ref{sec:meta}.

\emph{Mixing times.} The worst-case mixing time $t_{\rm mix}(1/4)=\min\{t:\max_x\|\delta_xA^t-\pi\|_{\rm TV}\le1/4\}$ \cite{LevinPeres2017} only requires one starting configuration per orbit, all of which are propagated at once with a batched matrix-free product.

Stopping rules and accuracy checks are given in Appendix~\ref{app:num}. Table~\ref{tab:summary} summarises the results at three interior points; full tables for $3\le L\le22$ are given in Appendix~\ref{app:tables}.

\begin{table}[t]\centering\small
\caption{Summary of numerical results at three interior points: stationary probability of a consensus configuration, second largest eigenvalue modulus, mean consensus time from the uniform law, Kac ratio $\E[\tau_c]\,\pi(\cH)$ and worst-case mixing time. Entries marked $^{*}$ are lower bounds obtained from candidate starting configurations; all others are exhaustive. Full tables are in Appendix~\ref{app:tables}.}\label{tab:summary}
\begin{tabular}{llllllc}\toprule
$(p,r)$ & $L$ & $\pi(0^L)$ & $|\lambda_2|$ & $\E[\tau_c]$ & $\E[\tau_c]\pi(\cH)$ & $t_{\rm mix}$\\\midrule
$(0.3,\,0.6)$ & 8 & $3.0604\times10^{-3}$ & 0.38603193 & $1.6398\times10^{2}$ & 1.00369 & 2\\
 & 16 & $9.3552\times10^{-6}$ & 0.38603227 & $5.3459\times10^{4}$ & 1.00023 & 2\\
 & 20 & $5.1739\times10^{-7}$ & 0.38603227 & $9.6643\times10^{5}$ & 1.00003 & 2$^{*}$\\\midrule
$(0.7,\,0.85)$ & 8 & $3.5115\times10^{-2}$ & 0.83188184 & $2.1274\times10^{1}$ & 1.49403 & 6\\
 & 16 & $1.2410\times10^{-3}$ & 0.83221539 & $4.5629\times10^{2}$ & 1.13249 & 8\\
 & 20 & $2.3292\times10^{-4}$ & 0.83221616 & $2.2775\times10^{3}$ & 1.06094 & 8$^{*}$\\\midrule
$(0.5,\,0.9)$ & 8 & $2.3102\times10^{-2}$ & 0.80000000 & $4.1193\times10^{1}$ & 1.90331 & 6\\
 & 16 & $5.4208\times10^{-4}$ & 0.80000000 & $1.1916\times10^{3}$ & 1.29191 & 8\\
 & 20 & $8.2718\times10^{-5}$ & 0.80000000 & $7.0578\times10^{3}$ & 1.16761 & 8$^{*}$\\
\bottomrule\end{tabular}
\end{table}

\begin{figure}[t]\centering
\includegraphics[width=\textwidth]{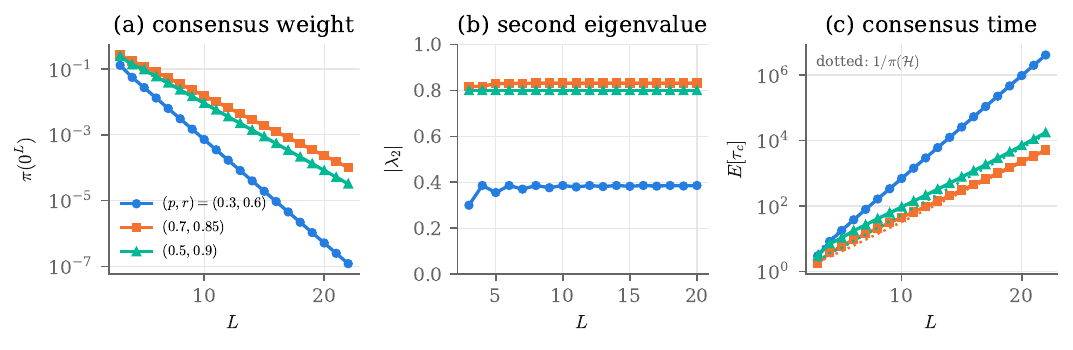}
\caption{Size dependence at three interior points. (a) Stationary probability of a consensus configuration. (b) Second largest eigenvalue modulus; at $p=1/2$ it equals $2r-1$ exactly (Proposition~\ref{prop:half}). (c) Mean consensus time from the uniform initial law (markers) and $1/\pi(\cH)$ (dotted).}\label{fig:sizes}
\end{figure}

Three features stand out (Table~\ref{tab:summary}, Figure~\ref{fig:sizes}). First, local observables and the spectral gap converge very fast with $L$: at $(0.3,0.6)$ the wall density is $0.519769289123\ldots$ for all $L\ge14$ and $|\lambda_2|$ is constant to ten digits from $L=12$. Second, the mixing time remains bounded or grows very slowly with $L$, consistent with an $O(\log L)$ behaviour; the worst start is a $0011$-periodic configuration at $(0.3,0.6)$ and the consensus state at $(0.5,0.9)$, where consensus is metastable. Third, consensus is exponentially rare: $\pi(0^L)\approx c\,\theta^L$ with $\theta\approx0.485$, $0.658$ and $0.625$ at the three points, and $\E[\tau_c]$ grows like $\theta^{-L}$, with $\rho(T)=1-1/\E[\tau_c]$ to high accuracy. The Kac ratio $\E[\tau_c]\,\pi(\cH)$ decreases towards one, the value expected when visits to $\cH$ become isolated rare events, since by Kac's lemma \cite{Kac1947} the mean return time to $\cH$ is $1/\pi(\cH)$. The approach is slower where visits come in clusters, as near the edge $r=1$. In opinion-dynamics language, any amount of hesitation in the interior of the square makes global consensus exponentially unlikely in large populations.

\section{Structure of the deterministic edges}\label{sec:edges}

By Proposition~\ref{prop:dual} the four edges reduce to two: the edge $r=1$ (standard contagion), equivalent to $r=0$ with $p\mapsto1-p$, and the edge $p=1$ (keep under ties), equivalent to $p=0$ with $r\mapsto1-r$.

\begin{proposition}[Edge $r=1$]\label{prop:edge-r1}
Let $r=1$ and $0<p<1$. The recurrent classes of the PCA are the two consensus states and, if and only if $L$ is even, the alternating two-cycle $a=0101\cdots01\leftrightarrow\bar a=1010\cdots10$. Every other configuration is transient.
\end{proposition}
\begin{proof}
At $r=1$ a cell whose neighbours agree copies them, and a cell whose neighbours disagree keeps its state with probability $p$ or flips it with probability $1-p$, both positive. Consensus is fixed, and in $a$ every cell sees two equal neighbours of the opposite value, so $a\mapsto\bar a\mapsto a$. For odd $L$ no configuration has all domains of length one. It remains to show that consensus is reachable from every other configuration. Call a maximal block of equal states a domain. If every tie cell keeps its state, cells in domains of length $\ge2$ keep their value, while every isolated cell adopts the common value of its two neighbours. A run of $m$ consecutive isolated cells between two domains of length $\ge2$ therefore shrinks by two cells per step (by one when $m=1$), so after finitely many steps all domains have length $\ge2$. If there are at least two domains, pick one, $D$, and in the next step flip only the tie cell adjacent to $D$ in the following domain. This enlarges $D$ by one cell; if the following domain was reduced to a single cell, that cell joins $D$ at the next step. Repeating, $D$ covers the ring. Every step used has positive probability.
\end{proof}

\begin{proposition}[Space-time sublattices at $r=1$, even $L$]\label{prop:stsub}
Let $r=1$ and let $L$ be even. Split space-time into the two sublattices $\{(i,t):i+t\ \text{even}\}$ and $\{(i,t):i+t\ \text{odd}\}$. If one of them is constant at time $t$, it stays constant, with the same value, at all later times. In particular, a chain started from a single defect on a consensus or on the alternating cycle is absorbed in one of the two traps on which the untouched sublattice keeps its value, never in the opposite consensus nor in the opposite phase of the alternating cycle.
\end{proposition}
\begin{proof}
The two neighbours of $(i,t+1)$ are $(i\pm1,t)$, which lie on the same sublattice as $(i,t+1)$. If that sublattice is constant at time $t$, the neighbours agree and the cell copies them (as $r=1$). For even $L$ the parity of $i+t$ is well defined on the ring. The last statement follows from Proposition~\ref{prop:edge-r1}.
\end{proof}

\begin{proposition}[Edge $p=1$]\label{prop:edge-p1}
Let $p=1$ and $0<r<1$. A configuration is absorbing if and only if every cell is a tie, $x_{i-1}\neq x_{i+1}$ for all $i$. Such configurations exist if and only if $4\mid L$, and they are the four rotations of $0011\,0011\cdots$. From a consensus state every configuration is reached in one step.
\end{proposition}
\begin{proof}
At $p=1$ tie cells keep their state, and a cell whose neighbours agree takes either value with positive probability. A configuration is fixed if and only if it has no agreeing cell. The condition $x_{i+1}=\bar x_{i-1}$ for all $i$ forces period four, which closes on the ring if and only if $4\mid L$. In a consensus state every cell has agreeing neighbours, so every configuration is reachable in one step.
\end{proof}

An exhaustive exact computation for $3\le L\le22$ (Appendix~\ref{app:verif-edge}) shows that, for these sizes, the recurrent states on the edge $p=1$ are exactly the frozen configurations when $4\mid L$, and otherwise form a single class containing both consensus states. We expect the same for every $L$.

\paragraph{Why parity and commensurability: stride-two cycles.} The condition for a frozen configuration at $p=1$, $x_{i+1}=\bar x_{i-1}$, says that the state alternates along each stride-two cycle, the same cycles that carry the Ising measure of Theorem~\ref{thm:ising}. An alternating pattern closes on a cycle if and only if the cycle has even length: for even $L$ the two cycles have length $L/2$, which gives $4\mid L$; for odd $L$ the single cycle has odd length and no frozen configuration exists. At $r=1$, the alternating two-cycle is the configuration that is constant on each stride-two cycle with opposite values on the two, which requires two cycles, i.e.\ even $L$; and the space-time sublattices of Proposition~\ref{prop:stsub} are the stride-two cycles followed in time. Hence the singular behaviour at the edges $p\in\{0,1\}$ is a commensurability effect present only when $4\mid L$, as in the case $L=4$ of \cite{MunozNuno2026}: the relaxation time at $p=1-\varepsilon$ diverges like $1/\varepsilon$ only when $4\mid L$ (Figure~\ref{fig:edges}c).

\section{Main result B: consensus times and metastability near the edges}\label{sec:meta}

\subsection{First-order metastable reduction}
Let $r=1-\varepsilon$. In every trap state ($0^L$, $1^L$ and, for even $L$, the two alternating configurations) all neighbourhoods are in agreement. Each cell therefore errs independently with probability $\varepsilon$, and to first order the trap is left through a single defect $y=P_0(s)\oplus e_i$, where $P_0$ is the deterministic motion on the traps ($0^L$ and $1^L$ fixed, the alternating states exchanged). From $y$ the edge chain is absorbed in the traps with probabilities $\Gamma(y,\cdot)$, resolved by phase for the alternating cycle. This is a PH absorption problem: with $T_1$ the transient part of the two-step edge chain $A(p,1)^2$ and $R_1$ its transitions into the traps, $\Gamma=(I-T_1)^{-1}R_1$. This yields the effective one-step map on the traps
\begin{equation}\label{eq:K}
K(\varepsilon)=P_0+\varepsilon R+O(\varepsilon^2),\qquad R(s,\cdot)=\sum_{i\in\Z_L}\Gamma\big(P_0(s)\oplus e_i,\cdot\big)-L\,e_{P_0(s)} .
\end{equation}
\begin{proposition}[First-order reduction]\label{prop:pert}
Let $0<p<1$. The eigenvalues of $A(p,1-\varepsilon)$ that tend to the unit circle as $\varepsilon\to0$ are $\omega+\varepsilon\nu+o(\varepsilon)$, where $\omega\in\{1,-1\}$ runs over the eigenvalues of $P_0$ and $\nu$ over the eigenvalues of the restriction of $R$ to the corresponding eigenspace of $P_0$. Writing $|\omega+\varepsilon\nu|=1-\varepsilon\mu_j+o(\varepsilon)$, $\mu_j=-\omega\,\mathrm{Re}\,\nu$, and letting $\mu(p,L)$ be the smallest nonzero $\mu_j$,
\begin{equation}\label{eq:tstar}
t^*(\varepsilon):=t_{\rm rel}=\frac{1}{1-|\lambda_2(A(p,1-\varepsilon))|}=\frac{1+o(1)}{\mu(p,L)\,\varepsilon},\qquad \varepsilon\to0,
\end{equation}
provided $\mu(p,L)>0$.
\end{proposition}
\begin{proof}
$\varepsilon\mapsto A(p,1-\varepsilon)$ is a polynomial matrix function. The eigenvalues $\pm1$ of the stochastic matrix $A(p,1)$ are semisimple, their left eigenvectors are supported on the traps, and the right eigenvectors are the (phase-resolved) absorption probabilities $\Gamma$. By first-order analytic perturbation theory for semisimple eigenvalues \cite{AvrachenkovFilarHowlett2013}, the perturbed eigenvalues are $\omega+\varepsilon\nu+o(\varepsilon)$, with $\nu$ the eigenvalues of $W\,A'(0)\,V$, where the rows of $W$ and the columns of $V$ are the left and right eigenvectors. The derivative $A'(0)$ restricted to the rows of a trap $s$ equals $\sum_i\delta_{P_0(s)\oplus e_i}-L\,\delta_{P_0(s)}$, since leaving the trap with $k$ defects has probability $\varepsilon^k(1-\varepsilon)^{L-k}$; hence $WA'(0)V$ is the restriction of $R$ in \eqref{eq:K}. All other eigenvalues of $A(p,1)$ have modulus $<1$ and stay away from the unit circle for small $\varepsilon$, which gives \eqref{eq:tstar}.
\end{proof}

This is the classical picture of nearly completely decomposable chains \cite{SimonAndo1961}: fast relaxation inside each basin of the edge chain and slow first-order exchange between traps. By duality, the same holds near $r=0$ with $\mu(1-p,L)$. Here $t^*$ is precisely the lifetime of the period-two oscillation that exists on the edge $r=0$ but not in the interior: the limits $t\to\infty$ and $r\to0$ do not commute, and \eqref{eq:tstar} quantifies the crossover that replaces the discontinuity observed in \cite{MunozNuno2026}. Consensus times behave accordingly: near $r=1$ consensus is a trap of lifetime of order $1/(\mu\varepsilon)$, which is why, at $(0.5,0.9)$, visits to $\cH$ come in clusters and the Kac ratio of Table~\ref{tab:summary} approaches one only slowly.

\subsection{The metastable rate in closed form}
Proposition~\ref{prop:stsub} fixes the structure of $R$. For odd $L$ there are only two traps, and $R=Lh\begin{psmallmatrix}-1&1\\1&-1\end{psmallmatrix}$, where $h$ is the probability that a single defect on $0^L$ is absorbed in $1^L$; thus $\mu=2Lh$. For even $L$, a defect on $0^L$ reaches the alternating cycle with probability $g_c$, in a phase fixed by the parity of the defect, and otherwise returns to $0^L$; a defect on the alternating cycle reaches a consensus with probability $g_a$, its value fixed by the untouched sublattice, and otherwise returns to the same phase. Rotation invariance distributes the $L$ defects equally between the two phases, so in the trap basis $(0^L,1^L,a,\bar a)$
\[
R=L\begin{pmatrix}-g_c&0&g_c/2&g_c/2\\0&-g_c&g_c/2&g_c/2\\g_a/2&g_a/2&0&-g_a\\g_a/2&g_a/2&-g_a&0\end{pmatrix}.
\]
On the eigenspace $\omega=1$ of $P_0$, spanned by $e_{0^L}$, $e_{1^L}$ and $e_a+e_{\bar a}$, the restriction of $R$ has eigenvalues $0$, $-Lg_c$ and $-L(g_c+g_a)$; on the eigenspace $\omega=-1$, spanned by $e_a-e_{\bar a}$, it is $Lg_a$. Hence $\mu=L\min(g_c,g_a)$. Write $q=1-p$.

\begin{theorem}[Metastable rate, even $L$]\label{thm:mu-even}
For even $L$ and $0<p<1$, $p\neq1/2$,
\begin{equation}\label{eq:mu-even}
g_c=\frac{q^{L-2}(q^2-p^2)}{q^L-p^L},\qquad g_a=\frac{p^{L-2}(q^2-p^2)}{q^L-p^L},\qquad \mu(p,L)=\frac{L\,(q^2-p^2)\,\min(p,q)^{L-2}}{q^L-p^L},
\end{equation}
and $\mu(1/2,L)=2$ by continuity.
\end{theorem}
\begin{proof}
Start from $0^L$ with a defect at cell $i_0$ at time $0$. By Proposition~\ref{prop:stsub} the untouched sublattice stays at $0$. A cell $(j,t+1)$ of the touched sublattice has both neighbours on the touched sublattice. If they agree it copies them; if they disagree, it keeps its own previous state, which lies on the untouched sublattice and equals $0$, with probability $p$, and becomes $1$ with probability $q$. On the touched sublattice, a ring of $n=L/2$ sites in the diagonal geometry, this is the Domany--Kinzel automaton \cite{DomanyKinzel1984} with $p_2=1$, $p_1=q$ and $p_0=0$: compact directed percolation \cite{Essam1989}. The active sites form an interval whose two ends move outwards with probability $q$ and inwards with probability $p$, independently. Its width therefore performs a lazy random walk on $\{0,\dots,n\}$ with steps $+1$, $-1$ and $0$ with probabilities $q^2$, $p^2$ and $2pq$. Width $0$ is the return to $0^L$, and width $n$ means that the touched sublattice is all $1$ while the untouched one is all $0$, that is, the alternating cycle. Gambler's ruin from width $1$ with ratio $\kappa=p^2/q^2$ gives $g_c=(1-\kappa)/(1-\kappa^{n})$, which is \eqref{eq:mu-even}. For a defect on the alternating cycle the untouched sublattice is constant (say $0$) and the touched one is all $1$ except the defect; a tie now keeps the untouched value $0$ with probability $p$, so the interval of $0$s grows with probability $p$ at each end. This is the same walk with $p\leftrightarrow q$, and gives $g_a$.
\end{proof}

\begin{remark}[Odd $L$]\label{rem:odd}
For odd $L$ the two space-time sublattices are joined into a single one, the percolation interval eventually interacts with itself, and the argument above does not apply. Exact computation suggests
\begin{equation}\label{eq:mu-odd}
h=\frac{q^{L-1}}{2\sum_{k=0}^{m}q^{2(m-k)}p^{2k}-p^{2m}},\qquad L=2m+1,
\end{equation}
which we have verified exactly for $L=3,5,7,9$ (Appendix~\ref{app:verif-odd}) and at three rational values of $p$ for $L=11,13$; we do not have a general proof.
\end{remark}

\begin{figure}[t]\centering
\includegraphics[width=\textwidth]{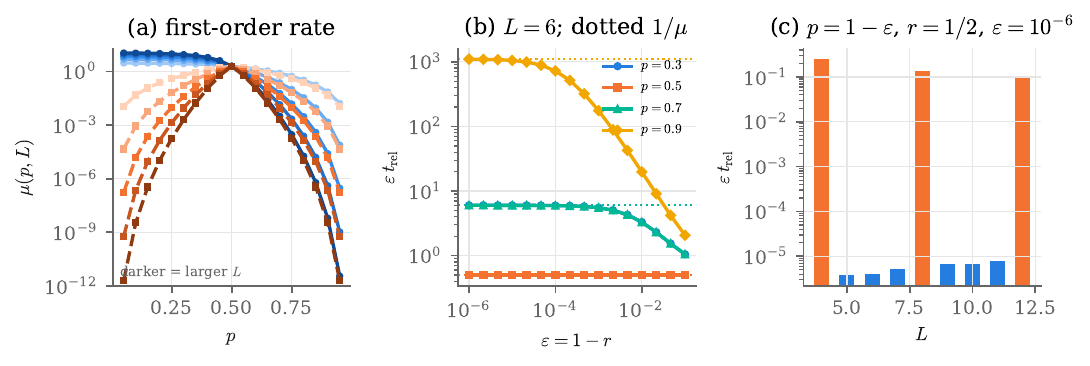}
\caption{Metastability near the edges. (a) First-order rate $\mu(p,L)$ for $L=3,5,\dots,11$ (blue, solid) and $L=4,6,\dots,12$ (orange, dashed); darker shades correspond to larger $L$. (b) $\varepsilon\,t_{\rm rel}$ for $L=6$ as $r=1-\varepsilon\to1$, with the prediction $1/\mu$ (dotted); the curves for $p=0.3$ and $p=0.7$ coincide by Proposition~\ref{prop:stag}. (c) Edge $p=1$: $\varepsilon\,t_{\rm rel}$ at $\varepsilon=10^{-6}$ remains finite only when $4\mid L$ (orange).}\label{fig:edges}
\end{figure}

The closed forms agree with the rates computed numerically through \eqref{eq:K} (Table~\ref{tab:mu}) and explain their behaviour. At $p=1/2$, $\mu=2$ for every $L$, in agreement with Proposition~\ref{prop:half}(b). For even $L$, $\mu(p)=\mu(1-p)$ is maximal at $p=1/2$ and decays like $L(1-(p/q)^2)(p/q)^{L-2}$ away from it, so the trap lifetime grows exponentially with the size. For odd $L$, $\mu\to L$ as $p\to0$ and $\mu\simeq2L(q/p)^{L-1}$ as $p\to1$. Figure~\ref{fig:edges}(b) compares \eqref{eq:tstar} with the exact spectral relaxation time: the agreement is excellent once $\varepsilon\ll\mu$, while for tiny $\mu$ second-order (two-defect) processes dominate at all accessible $\varepsilon$ and $t^*\propto\varepsilon^{-2}$ is observed instead.

\section{Infinite-volume behaviour}\label{sec:infinite}

On every finite ring the stationary distribution is unique in the interior and is a rational function of $(p,r)$, so there is no phase transition at finite $L$. The relevant question is whether the PCA on $\Z$ has several invariant measures, or a diverging correlation length or relaxation time, inside the square. For positive-rate PCA on $\Z$ this is delicate in general: the positive-rates conjecture fails for some one-dimensional PCA with large neighbourhoods \cite{Gacs2001}, and ergodicity is not known even for all binary PCA with two inputs \cite{Casse2022}. We do not prove ergodicity, but the results give a consistent picture.

\paragraph{Local structure theory.} The $L\to\infty$ limit of local observables can be computed directly on $\Z$ with the local structure theory of Gutowitz, Victor and Knight \cite{GVK1987}: an $n$-block measure is extended by its Bayesian (Markov of order $n-1$) extension, pushed through one step of the PCA, and iterated to a fixed point. This defines a spatial transfer matrix, in the spirit of transfer-matrix analyses of noisy majority automata \cite{LemoyMozeikaSeki2014}, and it is exact whenever the invariant measure is Markov of order $n-1$. We compute the push-forward with the same site-by-site contraction as in Section~\ref{sec:comp}. At $(0.3,0.6)$ the wall density is correct to $10^{-12}$ at order $n=8$ and agrees with the finite rings to thirteen digits. At $(0.7,0.85)$, where the rings converge slowly, the approximations converge geometrically in $n$ to $0.39383978451$, consistent with a geometric extrapolation of the ring values. On $p=1/2$ the approximation is exact from $n=3$, consistent with Corollary~\ref{cor:half}(d).

\paragraph{No sign of a transition in the interior.} On $p=1/2$ everything is explicit: correlations decay as $\xi^d$ with $\xi<1$, the gap $2\min(r,1-r)$ does not depend on $L$, all quantities are analytic in $r\in(0,1)$, and the only singularities are the zero-noise critical points \eqref{eq:diffusive} at the edges. Away from $p=1/2$, at the points of Table~\ref{tab:summary}, $|\lambda_2|$ converges with $L$ to a value strictly below one, $t_{\rm mix}$ is bounded or grows at most logarithmically, local observables converge geometrically both in $L$ and in the order of the local structure theory, and $\pi(0^L)$ decays at a well-defined exponential rate. This is what one expects from an ergodic PCA with exponentially decaying correlations. The non-analytic behaviour found in \cite{MunozNuno2026} is confined to the boundary.

\paragraph{The edges and the order of limits.} On the edges the gap closes in a size-dependent way. Near $r=1$, Theorem~\ref{thm:mu-even} gives a rate $\mu$ that is exponentially small in $L$ for even $L$ and $p\ne1/2$, so $t^*\simeq1/(\mu\varepsilon)$ grows exponentially with the size. The first-order description requires that traps are left through isolated defects, $L\varepsilon\ll1$, and that $\varepsilon\ll\mu$. The limits $\varepsilon\to0$ and $L\to\infty$ therefore do not commute: at fixed small $\varepsilon$ and large $L$ many defects coexist and the dynamics crosses over to coarsening of domains, which is beyond the present first-order theory. The parity and commensurability rules of Section~\ref{sec:edges} persist for arbitrarily large $L$ in the regime $L\varepsilon\ll1$.

\section{Discussion}\label{sec:disc}

We have shown that the PCA between rules 23, 77, 178 and 232 has an exact structure that goes well beyond the small rings of \cite{MunozNuno2026}. Its two main results, valid for every ring size, are the exact solution on the line $p=1/2$, whose stationary state is an Ising chain along the stride-two cycles and whose edges are zero-noise critical points with diffusive scaling, and the metastable time scale $t^*\simeq1/(\mu\varepsilon)$ near the edges, with $\mu$ given by PH absorption probabilities and, for even $L$, in closed form through compact directed percolation. The discontinuities that were the main qualitative finding of the first paper are thereby replaced by a quantitative crossover, together with parity ($L$ even or odd) and commensurability ($4\mid L$) rules for which edges are singular. A single combinatorial structure, the decomposition of the ring into stride-two cycles, underlies all of these. Exact symmetries, matrix-free linear algebra and PH methods make the rest of the square accessible up to $L=22$, and show rapid mixing and exponentially rare consensus in the interior.

Several directions remain open: (i) a proof of \eqref{eq:mu-odd} for every odd $L$, for which its form, a gambler's-ruin denominator corrected by one term, suggests a renewal argument for the self-interacting percolation interval; (ii) a proof of the recurrence structure on the edge $p=1$ for every $L$; (iii) ergodicity of the PCA on $\Z$ in the interior, at least in the attractive region $r\ge\max(p,1-p)$, and rigorous error bounds for the local structure theory; (iv) the coarsening regime at the edges when $L\varepsilon$ is not small; and (v) other matrix-analytic observables with an opinion-dynamics meaning, such as the statistics of opinion changes as a batch Markovian arrival process, which we will report separately.

\section*{CRediT authorship contribution statement}
\textbf{Francisco J. Mu\~noz:} Conceptualization, Methodology, Software, Validation, Formal analysis, Investigation, Writing -- original draft, Writing -- review \& editing. \textbf{Juan Carlos Nu\~no:} Conceptualization, Methodology, Software, Validation, Formal analysis, Investigation, Writing -- original draft, Writing -- review \& editing. Both authors contributed equally to this work.

\section*{Declaration of competing interest}
The authors declare that they have no known competing financial interests or personal relationships that could have appeared to influence the work reported in this paper.

\section*{Funding}
This research did not receive any specific grant from funding agencies in the public, commercial, or not-for-profit sectors.

\section*{Data availability}
All Python codes and data used in this work are available from the corresponding author upon request.

\section*{Declaration of generative AI and AI-assisted technologies in the manuscript preparation process}
During the preparation of this work the authors used Claude (Anthropic), under their supervision, for editing and revising the manuscript and for generating the figures. After using this tool, the authors reviewed and edited the content as needed and take full responsibility for the content of the published article.

\appendix

\section{Verifications by exact computation}\label{app:verif}
The statements below are established by exact computation (Python \texttt{fractions} module and integer bit operations, no rounding) for finitely many sizes; none of the results of Sections~\ref{sec:sym}--\ref{sec:meta} depends on them.

\subsection{Closed forms of \texorpdfstring{$\pi(p,r)$}{pi(p,r)} for \texorpdfstring{$L\le5$}{L<=5}}\label{app:verif-pi}
For fixed rational $r_j$, $\pi(x;p,r_j)$ is a rational function of $p$, reconstructed exactly from its values at many rational $p$; its coefficients are rational functions of $r$, reconstructed in the same way. To verify the result, write $\pi_O=N_O/D_O$ for the $n$ orbits, with degrees at most $d_p$ in $p$ and $d_r$ in $r$. Multiplying the stationarity equations of the lumped chain (whose entries have degree at most $L$ in each variable) and the normalisation by $\prod_OD_O$ gives polynomial identities of degree at most $nd_p+L$ in $p$ and $nd_r+L$ in $r$. We checked them in exact arithmetic on a grid of $(nd_p+L+1)\times(nd_r+L+1)$ rational points ($38\times38$ for $L=5$); a bivariate polynomial vanishing on such a grid is identically zero.

\subsection{Recurrence structure on the edge \texorpdfstring{$p=1$}{p=1}}\label{app:verif-edge}
A configuration $y$ is reachable from $x$ in one step if and only if $y$ coincides with $x$ on the tie cells of $x$. For each $3\le L\le22$ we computed, with integer operations on bit masks over all $2^L$ configurations, the backward closure of the set of consensus and frozen configurations under four positive-probability moves (agreeing cells set to $0$, set to $1$, copying, or anti-copying their neighbours). The closure is the whole configuration space. If $4\mid L$, frozen configurations are absorbing and reachable from consensus (Proposition~\ref{prop:edge-p1}), so they are the only recurrent states; otherwise consensus is reachable from every configuration, and the set reachable from consensus is the unique closed class. The same procedure reproduces Proposition~\ref{prop:edge-r1} for all $L\le22$.

\subsection{The odd-\texorpdfstring{$L$}{L} metastable rate}\label{app:verif-odd}
On the edge $r=1$ the entries of the lumped chain are polynomials of degree at most $L$ in $p$. With $n_t$ transient orbits, Cramer's rule gives $h=\det M_s/\det M$ with determinants of degree at most $n_tL$, so the right-hand side $N/D$ of \eqref{eq:mu-odd} satisfies $h=N/D$ identically if $\det M_s\,D-\det M\,N$ (degree at most $n_tL+L-1$) vanishes at $n_tL+L$ points. We checked this in exact arithmetic at $9$, $35$, $119$ and $405$ rational points for $L=3,5,7,9$.

\section{Numerical methods and tolerances}\label{app:num}
All computations were implemented in Python, using NumPy and SciPy for the numerical linear algebra, in IEEE double precision. Stationary distributions: power iteration stopped when $\|v_{k+1}-v_k\|_1<10^{-13}$, with residuals $\|\pi A-\pi\|_1<7\times10^{-14}$. Leading eigenvalues: implicitly restarted Arnoldi (ARPACK through SciPy \cite{ARPACK}) with tolerance $10^{-9}$; relative residuals are at most $2.4\times10^{-10}$, the difference with dense diagonalisation for $L=12$ is at most $3\times10^{-12}$, and at $p=1/2$ the computed $|\lambda_2|$ equals $2r-1$ to $6\times10^{-15}$. PH means: survival ratios propagated to a relative tolerance $10^{-12}(1-\rho)$ plus a geometric tail; for $L\le8$ they agree with the exact lumped values to $12$ significant digits. The matrix-free products agree with the dense matrix to $10^{-15}$ for $3\le L\le9$.

\section{Exact results and closed forms for small rings}\label{app:exact}
Table~\ref{tab:exact} lists exact results on the lumped chain at the rational point $(p,r)=(3/10,3/5)$, obtained with exact rational arithmetic (Section~\ref{sec:comp}). The values for $L=3,4$ coincide with the closed forms of \cite{MunozNuno2026}.

\begin{table}[!htbp]\centering\small
\caption{Exact results at $(p,r)=(3/10,3/5)$: number of $G$-orbits, stationary probability of one consensus configuration, and mean consensus time from the uniform initial law. Exact fractions are shown when short; the decimal columns are truncations of exact rationals.}\label{tab:exact}
\begin{tabular}{rrllll}\toprule
$L$ & orbits & $\pi(0^L)$ exact & decimal & $\E[\tau_c]$ exact & decimal\\\midrule
3 & 2 & $25/194$ & 0.1288659794 & $3$ & 3.0000000000 \\
4 & 4 & $8281/149300$ & 0.0554655057 & $1650325/198744$ & 8.3037726925 \\
5 & 4 & $1784497/65965634$ & 0.0270519192 & $96208625/5353491$ & 17.9711939368 \\
6 & 8 & -- & 0.0130834171 & -- & 37.9820441112 \\
7 & 9 & -- & 0.0063099361 & -- & 79.3743447798 \\
8 & 18 & -- & 0.0030603837 & -- & 163.9803518612 \\
\bottomrule\end{tabular}
\end{table}

\subsection*{Closed-form stationary distributions for \mbox{$L=3,4,5$}}
For $L=3,4,5$ and $0<p,r<1$ the stationary distribution is given below, one configuration per orbit of $G=D_L\times\Z_2$; every configuration of an orbit has the same probability. For each $L$ the denominator is common to all orbits, and polynomials are written in powers of $p$ with coefficients in $r$. The formulas were obtained and verified as explained in Appendix~\ref{app:verif-pi}. As a check, at $(p,r)=(3/10,3/5)$ they reproduce Table~\ref{tab:exact}, and the probabilities of all $2^L$ configurations add up to one.

\paragraph{$L=3$.} $\pi(x)=N_x/D_3$ with
{\small\begin{align*}
D_{3}&=\bigl(2+4r-6r^{2}\bigr)\\
&\quad+\bigl(-4+4r\bigr)p\\
&\quad+2p^{2}\\[3pt]
N_{000}&=\bigl(1-r\bigr)\\
&\quad+\bigl(-2+2r\bigr)p\\
&\quad+p^{2}\\[3pt]
N_{001}&=\bigl(r-r^{2}\bigr)
\end{align*}}
\paragraph{$L=4$.} $\pi(x)=N_x/D_4$ with
{\small\begin{align*}
D_{4}&=\bigl(-4r^{2}+16r^{3}-28r^{4}+24r^{5}-8r^{6}\bigr)\\
&\quad+\bigl(-4-8r+16r^{2}-16r^{3}+8r^{4}\bigr)p\\
&\quad+\bigl(16-8r^{2}+16r^{3}-8r^{4}\bigr)p^{2}\\
&\quad+\bigl(-32+16r-16r^{2}\bigr)p^{3}\\
&\quad+\bigl(36-8r+8r^{2}\bigr)p^{4}\\
&\quad-24p^{5}\\
&\quad+8p^{6}\\[3pt]
N_{0000}&=\bigl(-2+5r-5r^{2}+2r^{3}\bigr)p\\
&\quad+\bigl(8-19r+17r^{2}-6r^{3}\bigr)p^{2}\\
&\quad+\bigl(-14+26r-18r^{2}+4r^{3}\bigr)p^{3}\\
&\quad+\bigl(14-16r+6r^{2}\bigr)p^{4}\\
&\quad+\bigl(-8+4r\bigr)p^{5}\\
&\quad+2p^{6}\\[3pt]
N_{0001}&=\bigl(-2r+4r^{2}-4r^{3}+2r^{4}\bigr)p\\
&\quad+\bigl(4r-6r^{2}+4r^{3}-2r^{4}\bigr)p^{2}\\
&\quad+\bigl(-4r+4r^{2}\bigr)p^{3}\\
&\quad+\bigl(2r-2r^{2}\bigr)p^{4}\\[3pt]
N_{0011}&=\bigl(-r^{2}+4r^{3}-7r^{4}+6r^{5}-2r^{6}\bigr)\\
&\quad+\bigl(-2r^{2}+4r^{3}-2r^{4}\bigr)p\\
&\quad+\bigl(2r^{2}-4r^{3}+2r^{4}\bigr)p^{2}\\[3pt]
N_{0101}&=\bigl(-r+r^{2}-2r^{3}\bigr)p\\
&\quad+\bigl(3r-r^{2}+6r^{3}\bigr)p^{2}\\
&\quad+\bigl(-2-2r-6r^{2}-4r^{3}\bigr)p^{3}\\
&\quad+\bigl(4+4r+6r^{2}\bigr)p^{4}\\
&\quad+\bigl(-4-4r\bigr)p^{5}\\
&\quad+2p^{6}
\end{align*}}
\paragraph{$L=5$.} $\pi(x)=N_x/D_5$ with
{\small\begin{align*}
D_{5}&=\bigl(-2r-10r^{2}+16r^{3}+50r^{4}-228r^{5}+344r^{6}-220r^{7}+50r^{8}\bigr)\\
&\quad+\bigl(-4+4r-16r^{2}-32r^{3}+180r^{4}-136r^{5}-36r^{6}+40r^{7}\bigr)p\\
&\quad+\bigl(20-36r+52r^{2}+128r^{3}-480r^{4}+420r^{5}-104r^{6}\bigr)p^{2}\\
&\quad+\bigl(-44+72r-144r^{2}+128r^{3}+60r^{4}-72r^{5}\bigr)p^{3}\\
&\quad+\bigl(58-76r+136r^{2}-180r^{3}+60r^{4}\bigr)p^{4}\\
&\quad+\bigl(-52+40r-12r^{2}+24r^{3}\bigr)p^{5}\\
&\quad+\bigl(32-20r-8r^{2}\bigr)p^{6}\\
&\quad+\bigl(-12+8r\bigr)p^{7}\\
&\quad+2p^{8}\\[3pt]
N_{00000}&=\bigl(-r+5r^{2}-12r^{3}+15r^{4}-9r^{5}+2r^{6}\bigr)\\
&\quad+\bigl(-2+12r-38r^{2}+74r^{3}-80r^{4}+42r^{5}-8r^{6}\bigr)p\\
&\quad+\bigl(10-48r+116r^{2}-176r^{3}+150r^{4}-60r^{5}+8r^{6}\bigr)p^{2}\\
&\quad+\bigl(-22+86r-162r^{2}+184r^{3}-110r^{4}+24r^{5}\bigr)p^{3}\\
&\quad+\bigl(29-83r+108r^{2}-80r^{3}+25r^{4}\bigr)p^{4}\\
&\quad+\bigl(-26+50r-36r^{2}+12r^{3}\bigr)p^{5}\\
&\quad+\bigl(16-20r+6r^{2}\bigr)p^{6}\\
&\quad+\bigl(-6+4r\bigr)p^{7}\\
&\quad+p^{8}\\[3pt]
N_{00001}&=\bigl(-r^{2}+4r^{3}-7r^{4}+5r^{5}-r^{7}\bigr)\\
&\quad+\bigl(-2r+8r^{2}-20r^{3}+30r^{4}-20r^{5}+2r^{6}+2r^{7}\bigr)p\\
&\quad+\bigl(6r-20r^{2}+42r^{3}-58r^{4}+39r^{5}-9r^{6}\bigr)p^{2}\\
&\quad+\bigl(-8r+16r^{2}-16r^{3}+16r^{4}-8r^{5}\bigr)p^{3}\\
&\quad+\bigl(7r-10r^{2}+r^{3}+2r^{4}\bigr)p^{4}\\
&\quad+\bigl(-4r+6r^{2}-2r^{3}\bigr)p^{5}\\
&\quad+\bigl(r-r^{2}\bigr)p^{6}\\[3pt]
N_{00011}&=\bigl(-r^{2}+2r^{3}+2r^{4}-14r^{5}+24r^{6}-18r^{7}+5r^{8}\bigr)\\
&\quad+\bigl(-6r^{3}+24r^{4}-34r^{5}+20r^{6}-4r^{7}\bigr)p\\
&\quad+\bigl(3r^{2}-6r^{3}-3r^{4}+12r^{5}-6r^{6}\bigr)p^{2}\\
&\quad+\bigl(-4r^{2}+12r^{3}-12r^{4}+4r^{5}\bigr)p^{3}\\
&\quad+\bigl(r^{2}-2r^{3}+r^{4}\bigr)p^{4}\\[3pt]
N_{00101}&=\bigl(-2r^{3}+7r^{4}-12r^{5}+10r^{6}-3r^{7}\bigr)\\
&\quad+\bigl(-2r^{2}+8r^{3}-20r^{4}+32r^{5}-24r^{6}+6r^{7}\bigr)p\\
&\quad+\bigl(-r^{2}+12r^{3}-17r^{4}+3r^{5}+3r^{6}\bigr)p^{2}\\
&\quad+\bigl(-2r+6r^{2}-20r^{3}+24r^{4}-8r^{5}\bigr)p^{3}\\
&\quad+\bigl(2r+r^{2}-r^{3}-2r^{4}\bigr)p^{4}\\
&\quad+\bigl(-2r+2r^{3}\bigr)p^{5}\\
&\quad+\bigl(r-r^{2}\bigr)p^{6}
\end{align*}}

\clearpage
\section{Full numerical tables}\label{app:tables}
Tables~\ref{tab:num}--\ref{tab:lst} give the complete numerical results summarised in the main text.

\begin{table}[!htbp]\centering\small
\caption{Double-precision results for larger rings. Wall: stationary density of domain walls. $\E[\tau_c]$: mean consensus time from the uniform law. The last column is $\E[\tau_c]\,\pi(\cH)$. Eigenvalues were computed for $L\le 20$.}\label{tab:num}
\begin{tabular}{llllllr}\toprule
$(p,r)$ & $L$ & $\pi(0^L)$ & wall & $|\lambda_2|$ & $\E[\tau_c]$ & $\E[\tau_c]\pi(\cH)$\\\midrule
$(0.3,\,0.6)$ & 8 & $3.0604\times10^{-3}$ & 0.51976949 & 0.38603193 & $1.6398\times10^{2}$ & 1.00369 \\
 & 12 & $1.6916\times10^{-4}$ & 0.51976929 & 0.38603227 & $2.9599\times10^{3}$ & 1.00138 \\
 & 16 & $9.3552\times10^{-6}$ & 0.51976929 & 0.38603227 & $5.3459\times10^{4}$ & 1.00023 \\
 & 20 & $5.1739\times10^{-7}$ & 0.51976929 & 0.38603227 & $9.6643\times10^{5}$ & 1.00003 \\
 & 21 & $2.5090\times10^{-7}$ & 0.51976929 & -- & $1.9928\times10^{6}$ & 1.00002 \\
 & 22 & $1.2167\times10^{-7}$ & 0.51976929 & -- & $4.1094\times10^{6}$ & 1.00001 \\
\midrule
$(0.7,\,0.85)$ & 8 & $3.5115\times10^{-2}$ & 0.39220300 & 0.83188184 & $2.1274\times10^{1}$ & 1.49403 \\
 & 12 & $6.6107\times10^{-3}$ & 0.39373810 & 0.83220117 & $9.6780\times10^{1}$ & 1.27957 \\
 & 16 & $1.2410\times10^{-3}$ & 0.39383384 & 0.83221539 & $4.5629\times10^{2}$ & 1.13249 \\
 & 20 & $2.3292\times10^{-4}$ & 0.39383944 & 0.83221616 & $2.2775\times10^{3}$ & 1.06094 \\
 & 21 & $1.5331\times10^{-4}$ & 0.39383961 & -- & $3.4253\times10^{3}$ & 1.05026 \\
 & 22 & $1.0091\times10^{-4}$ & 0.39383970 & -- & $5.1605\times10^{3}$ & 1.04149 \\
\midrule
$(0.5,\,0.9)$ & 8 & $2.3102\times10^{-2}$ & 0.50000000 & 0.80000000 & $4.1193\times10^{1}$ & 1.90331 \\
 & 12 & $3.5510\times10^{-3}$ & 0.50000000 & 0.80000000 & $2.1480\times10^{2}$ & 1.52549 \\
 & 16 & $5.4208\times10^{-4}$ & 0.50000000 & 0.80000000 & $1.1916\times10^{3}$ & 1.29191 \\
 & 20 & $8.2718\times10^{-5}$ & 0.50000000 & 0.80000000 & $7.0578\times10^{3}$ & 1.16761 \\
 & 21 & $5.1699\times10^{-5}$ & 0.49999940 & -- & $1.1091\times10^{4}$ & 1.14683 \\
 & 22 & $3.2312\times10^{-5}$ & 0.50000000 & -- & $1.7469\times10^{4}$ & 1.12893 \\
\bottomrule\end{tabular}
\end{table}

\begin{table}[!htbp]\centering\small
\caption{Worst-case mixing time $t_{\rm mix}(1/4)$. Exhaustive over all orbits except entries marked $^{*}$, which use candidate starting configurations (consensus, alternating, $0011$-periodic and half-and-half) and are lower bounds.}\label{tab:tmix}
\begin{tabular}{rccc}\toprule
$L$ & $(0.3,0.6)$ & $(0.7,0.85)$ & $(0.5,0.9)$\\\midrule
3 & 1 & 4 & 4 \\
4 & 2 & 4 & 5 \\
5 & 2 & 5 & 5 \\
6 & 2 & 5 & 6 \\
7 & 2 & 6 & 6 \\
8 & 2 & 6 & 6 \\
9 & 2 & 6 & 6 \\
10 & 2 & 7 & 7 \\
11 & 2 & 7 & 7 \\
12 & 2 & 7 & 7 \\
13 & 2 & 7 & 7 \\
14 & 2 & 7 & 7 \\
15 & 2 & 7 & 8 \\
16 & 2 & 8 & 8 \\
17 & 2 & 8 & 8$^{*}$ \\
18 & 2$^{*}$ & 8$^{*}$ & 8$^{*}$ \\
19 & 2$^{*}$ & 8$^{*}$ & 8$^{*}$ \\
20 & 2$^{*}$ & 8$^{*}$ & 8$^{*}$ \\
\bottomrule\end{tabular}
\end{table}

\begin{table}[!htbp]\centering\small
\caption{First-order metastable rate $\mu(p,L)$ near the edge $r=1$, computed numerically from phase-type absorption probabilities through \eqref{eq:K}. It coincides with Theorem~\ref{thm:mu-even} for even $L$ and with \eqref{eq:mu-odd} for odd $L$. $t^*\simeq1/(\mu\varepsilon)$; by duality, $\mu(1-p,L)$ governs the approach to $r=0$.}\label{tab:mu}
\begin{tabular}{rccccc}\toprule
$L$ & $p=0.1$ & $0.3$ & $0.5$ & $0.7$ & $0.9$\\\midrule
3 & 2.98160 & 2.74766 & 2.00000 & 0.80597 & 0.07229 \\
4 & 0.04878 & 0.62069 & 2.00000 & 0.62069 & 0.04878 \\
5 & 4.93865 & 4.16479 & 2.00000 & 0.23512 & 0.00149 \\
6 & $9.032\times10^{-4}$ & 0.16627 & 2.00000 & 0.16627 & $9.032\times10^{-4}$ \\
7 & 6.91359 & 5.73532 & 2.00000 & 0.05994 & $2.570\times10^{-5}$ \\
8 & $1.487\times10^{-5}$ & 0.04051 & 2.00000 & 0.04051 & $1.487\times10^{-5}$ \\
9 & 8.88889 & 7.35189 & 2.00000 & 0.01413 & $4.080\times10^{-7}$ \\
10 & $2.294\times10^{-7}$ & 0.00929 & 2.00000 & 0.00929 & $2.294\times10^{-7}$ \\
11 & 10.86420 & 8.98070 & 2.00000 & 0.00317 & $6.156\times10^{-9}$ \\
12 & $3.399\times10^{-9}$ & 0.00205 & 2.00000 & 0.00205 & $3.399\times10^{-9}$ \\
\bottomrule\end{tabular}
\end{table}

\begin{table}[!htbp]\centering\small
\caption{$\varepsilon\,t_{\rm rel}$ at $p=1-\varepsilon$, $r=1/2$, $\varepsilon=10^{-6}$. The relaxation time diverges as $\varepsilon\to0$ only when $4\mid L$.}\label{tab:pedge}
\begin{tabular}{rcc}\toprule
$L$ & $\varepsilon\,t_{\rm rel}$ & $4\mid L$\\\midrule
4 & $0.250$ & yes \\
5 & $3.81\times10^{-6}$ & no \\
6 & $4.00\times10^{-6}$ & no \\
7 & $5.32\times10^{-6}$ & no \\
8 & $0.137$ & yes \\
9 & $6.73\times10^{-6}$ & no \\
10 & $6.67\times10^{-6}$ & no \\
11 & $7.94\times10^{-6}$ & no \\
12 & $0.093$ & yes \\
\bottomrule\end{tabular}
\end{table}

\begin{table}[!htbp]\centering\small
\caption{Local structure theory of order $n$: stationary wall density and $C_2=\E[\sigma_0\sigma_2]$ on $\mathbb{Z}$, compared with finite rings (bottom).}\label{tab:lst}
\begin{tabular}{lcccc}\toprule
& \multicolumn{2}{c}{$(p,r)=(0.3,0.6)$} & \multicolumn{2}{c}{$(0.7,0.85)$}\\
$n$ & wall & $C_2$ & wall & $C_2$\\\midrule
2 & 0.5197725443054 & -- & 0.3991538229443 & -- \\
3 & 0.5197884686758 & 0.0105809971453 & 0.3950286361692 & 0.1771064876815 \\
4 & 0.5197705453827 & 0.0105057478968 & 0.3939534823578 & 0.1786866256240 \\
5 & 0.5197693493102 & 0.0104793702364 & 0.3938433902695 & 0.1802769438763 \\
6 & 0.5197692876672 & 0.0104792309169 & 0.3938412182703 & 0.1802908072717 \\
8 & 0.5197692891235 & 0.0104792474899 & 0.3938397982736 & 0.1803042427042 \\
10 & 0.5197692891234 & 0.0104792474895 & 0.3938397845830 & 0.1803043453744 \\
12 & 0.5197692891234 & 0.0104792474895 & 0.3938397845109 & 0.1803043459185 \\
14 & 0.5197692891234 & 0.0104792474895 & 0.3938397845121 & 0.1803043459141 \\
\midrule
ring $L=16$ & 0.5197692891234 & & 0.3938338351640 & \\
ring $L=20$ & 0.5197692891234 & & 0.3938394391272 & \\
ring $L=22$ & 0.5197692891234 & & 0.3938397013370 & \\
\bottomrule\end{tabular}
\end{table}

\end{document}